\documentclass[
  reprint,
  aps,
  prl,
  superscriptaddress,
  longbibliography
]{revtex4-2}

\usepackage{amsmath,amssymb,bm}
\usepackage{graphicx}
\usepackage{siunitx}
\usepackage{xcolor}
\usepackage{hyperref}
\usepackage{color,ulem}
\usepackage{soul}

\begin{document}

\title{Thermalization and dephasing in an isolated system of coupled qubits}
\author{Jukka P. Pekola}
\affiliation{Pico group, Department of Applied Physics, Aalto University School of Science, P.O. Box 13500, 00076 Aalto, Finland}
\author{Bayan Karimi}
\affiliation{Pico group, Department of Applied Physics, Aalto University School of Science, P.O. Box 13500, 00076 Aalto, Finland}
\affiliation{Pritzker School of Molecular Engineering, University of Chicago, Chicago IL 60637, USA}
\date{\today}

\begin{abstract}
A system evolving unitarily does not dephase or thermalize by definition. Yet, if one considers a part of a unitary system, situation is different. Here we analyze an isolated set of coupled qubits taking one qubit at a time as a subsystem, and writing self-consistently a weak-coupling master equation for it. Weakly anharmonic oscillators and ideal qubits (two-level systems) demonstrate dephasing, i.e. the diagonal elements of the density matrix reach a steady-state in the long-time limit, and the off-diagonal elements of ideal qubits decay exponentially. For the linear coupling between the qubits, only the degenerate ones interact, and the system does not reach thermal distribution. However, non-linearity, in our analysis in form of three-wave mixing, ensures a thermal distribution in the long time limit, where the temperature is determined uniquely by the energy in the initial non-thermal state. Finally, we present a potential experimental scheme based on a superconducting quantum circuit.
\end{abstract}

\maketitle
{\sl Introduction:} The interplay between quantum unitary evolution and thermalization has attracted theoretical attention for a long time, and more recently, experimental interest as well. This is enabled by rapid advances in quantum technology, which have allowed synthetic quantum systems, such as superconducting quantum circuits and cold atoms, to increasingly realize nearly unitary evolution \cite{Ueda2018,Abanin2019,Rigol2016,Nandkishore2015,Andersen2025,Guo2026,Weiss2006,Schmiedmayer2012,Kaufman2016,Neill2016,Chen2021,Higginbotham2026}. Quantum information processing relies on unitary evolution, whereas thermalization to Gibbs state with vanishing off-diagonal elements in a density matrix corresponds to return to classical dynamics and loss of potential quantum advantage. An open quantum system coupled to a classical heat bath undergoes relaxation and dephasing, eventually approaching thermal equilibrium. Understanding and controlling these processes is a central challenge in practical quantum engineering and has been extensively studied. On the contrary, much less is understood about internal thermalization of a closed (isolated) quantum system, which is the topic of this work.
\begin{figure}
		\centering
		\includegraphics [width=\columnwidth] {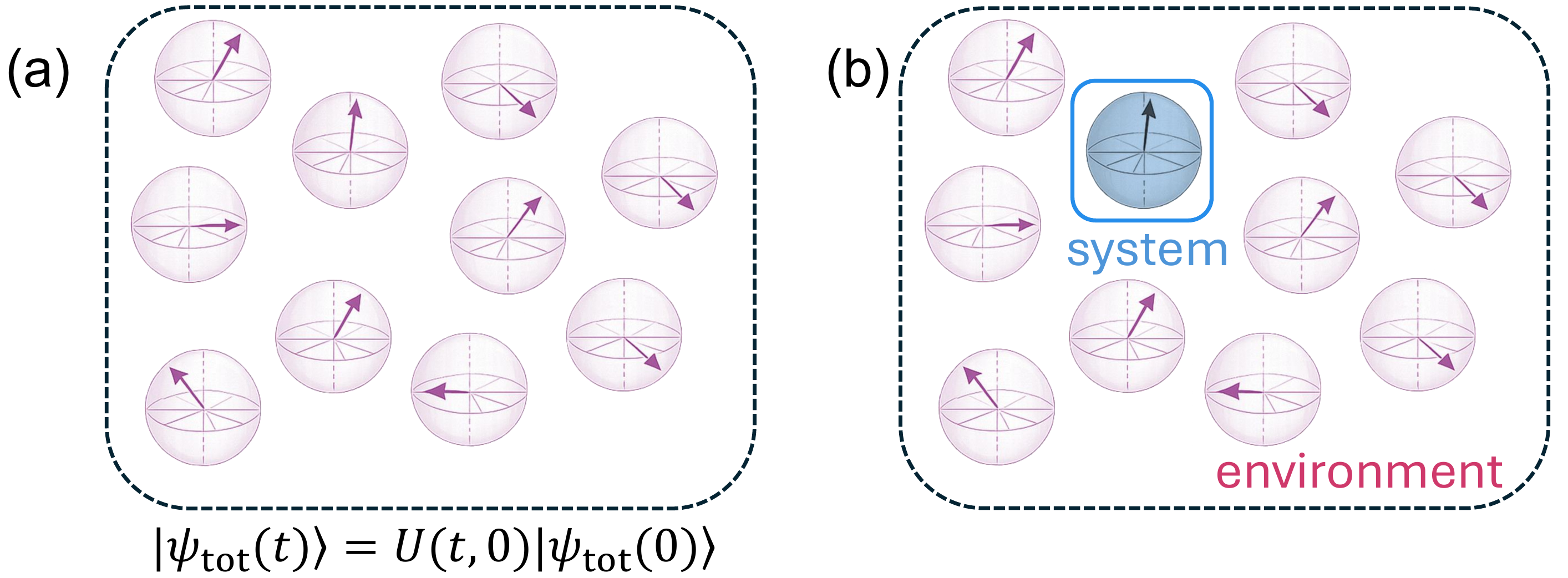}
		\caption{An isolated ensemble of coupled qubits. (a) Globally the ensemble evolves unitarily. (b) Here we study the ``system" formed in this case of a single qubit, with the rest of the isolated ensemble forming the bath. 
			\label{qubit+environment}}
\end{figure}

We consider an archetypal isolated quantum system, a set of $N$ all-to-all coupled qubits, either in form of weakly anharmonic oscillators (WAO) or ideal two-level qubits, which can be realized these days on solid-state platforms to a good approximation~\cite{XuntaoWu}. To study the internal dynamics of this ensemble, we treat a single qubit as the system and the remaining ones as its effective bath. Using a weak-coupling master equation, we perform and repeat the analysis self-consistently across all qubits. We consider two coupling mechanisms: cubic non-linearity, i.e. three-wave mixing and linear one-to-one coupled set of qubits conserving the number of excitations (see End Matter). In the nonlinear coupling, the system reaches a thermal state under all cases we consider, unlike a linearly coupled system. %We illuminate thermalization by analyzing the dynamics of a qubit in the ensemble formed by the rest of them. We also revisit some exact results on a linearly coupled system of qubits outside the weak-coupling regime, demonstrating that the absence of thermalization in a linear system is not a special property of extreme weak coupling. 
We also illustrate the emergence of dephasing of a system (single qubit) based on the same master equation analysis: this happens both with linear and non-linear coupling, but for linear coupling only in the large $N$ limit.

{\sl Model:} First, we consider a generic isolated, all-to-all-coupled multi-qubit system, as illustrated in Fig.~\ref{qubit+environment}. The $j$:th qubit has an energy splitting between its excited and ground states of $\hbar\Omega_j$. We denote its coupling to the remaining $N-1$ qubits by $\hat{\mathbb{V}}^{(j)}$. Our approach is to treat these $N-1$ qubits as an effective bath for qubit $j$, and apply the same construction self-consistently to each qubit, $j=1,\ldots,N$. We then write the standard weak-coupling master equation for the density matrix $\rho^{(j)}$ of the $j$:th qubit in the interaction picture as
\begin{equation}\label{eq1}
	\dot{\rho}^{(j)}(t)=-\frac{1}{\hbar^2}{\rm Tr}_{\rm B} \int_{-\infty}^{t}dt' [[\rho^{(j)}(t')\rho_{\rm B},\hat{\mathbb{V}}^{(j)}_I(t')],\hat{\mathbb{V}}^{(j)}_I(t)],
\end{equation}
with the coupling Hamiltonian also in the interaction picture as $\hat{\mathbb{V}}^{(j)}_I(t)=e^{i\hat H_0t/\hbar}\hat{\mathbb{V}}^{(j)}e^{-i\hat H_0t/\hbar}$, with $\hat H_0$ the unperturbed Hamiltonian of the system in the absence of couplings, and $\hat{\mathbb{V}}^{(j)}$ expressed in the Schr\"odinger picture. ${\rm Tr}_{\rm B}$ represents trace over the bath.

To start with, our qubit is a multilevel system with unequal spacing between levels (WAO). As an example, this is the case of the nowadays most common superconducting qubit, transmon \cite{Koch2007}. We will treat the anharmonicity perturbatively and take into account all the levels $n_k=0,1,2,...,\infty$ for the $k$:th oscillator. 
\begin{figure}
		\centering
		\includegraphics [width=\columnwidth] {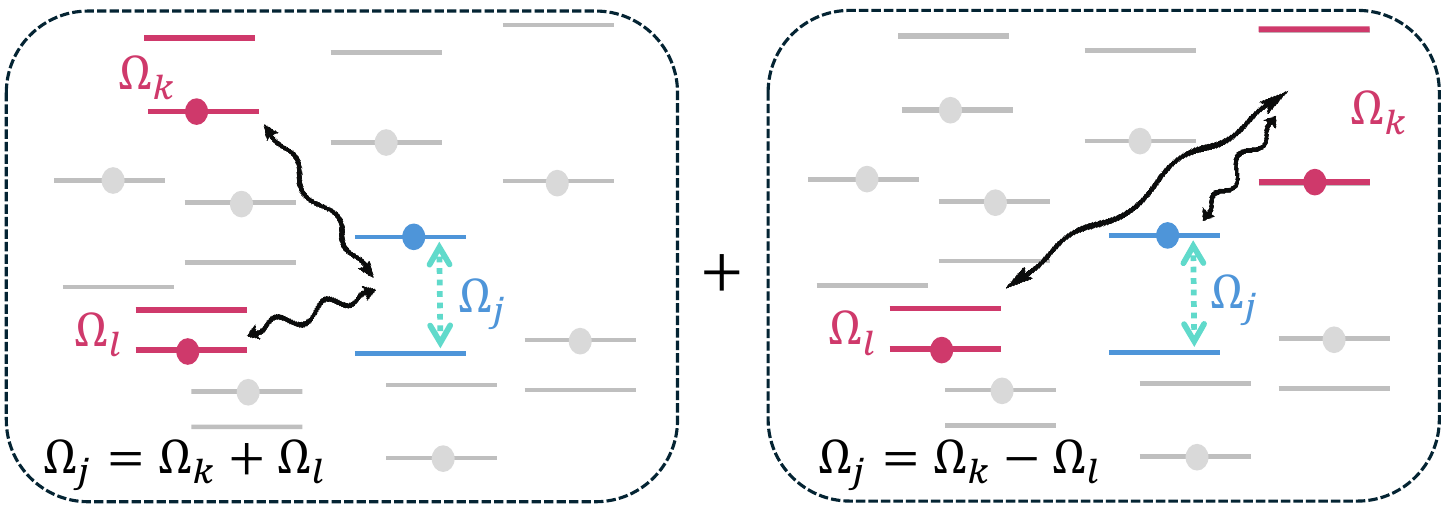}
		\caption{Three-wave mixing in all-to-all coupled qubits. Unlike one-to-one coupling, three-wave mixing allows interaction and energy exchange between non-degenerate qubits as well.
			\label{linvs3w}}
\end{figure}

In three-wave mixing~\cite{Roch2012,Frattini2017,Ganesan2017,Visa2018,Roch2022,Pla2022,Manucharyan2023,Buccheri2026,3wave-PK,Ankerhold2026}, see Fig. \ref{linvs3w}, the number of excitations is not conserved, and due to the conditions of energy conservation among three qubits in an emission/absorption process, there is room for relaxation. It warrants distribution of population among all qubits and thus forms an ergodic system. To find the way the $j$:th qubit couples to the ``bath" ones, we first note that the total Hamiltonian of the qubit system in case of pure three-wave mixing as perturbation can be written as 
\begin{equation} \label{eq6}
H=H_0+\sum_{p,q,r} (M_{pqr}a_p^\dagger a_qa_r+M_{pqr}^*a_r^\dagger a_q^\dagger a_p), 
\end{equation}
where $M_{pqr}$ are the coupling parameters, and the qubit $l$ is excited (annihilated) by the ladder operators $a_l^\dagger$ ($a_l$). The terms in the sum that involve coupling to the $j$:th qubit can be identified by setting either $p,q$ or $r$ equal to $j$, the rest of the coupling terms remain internal ones in the bath. In this case,
\begin{eqnarray}\label{eq6a}	
&&\hat{\mathbb{V}}^{(j)}=\sum_{k,\ell=1}^{N}\big{(}M_{jk\ell}\hat{a}_j^\dagger \hat{a}_k \hat{a}_\ell+M_{jk\ell}^*\hat{a}_\ell^\dagger \hat{a}_k^\dagger \hat{a}_j\nonumber\\&&~~~~~~~~~~~~~~~+2M_{k\ell j}\hat{a}_k^\dagger\hat{a}_\ell\hat{a}_j+2M_{k\ell j}^*\hat{a}_j^\dagger \hat{a}_\ell^\dagger \hat{a}_k\big{)}.
\end{eqnarray}
The factor $2$ in the last two terms arises as letting either $q$ or $r$ be equal to $j$ in Eq. \eqref{eq6} yields two identical contributions, since the operators for distinguishable qubits commute. 
By standard manipulations we can combine Eqs. \eqref{eq1} and \eqref{eq6a} and obtain the master equation for the diagonal elements of $\rho^{(j)}$ on the level $n_j$ in the harmonic oscillator approximation in the form
\begin{eqnarray} \label{eq5duplicate}
&&\dot{\rho}_{n_j,n_j}^{(j)}=-\Gamma_{0,n_j}^{(j)}\rho_{n_j,n_j}^{(j)}\nonumber\\&&~~~~~~~~~~~~+\Gamma_{\downarrow,n_j}^{(j)}\rho_{n_j+1,n_j+1}^{(j)}+\Gamma_{\uparrow,n_j}^{(j)}\rho_{n_j-1,n_j-1}^{(j)},
\end{eqnarray}
where we have
\begin{widetext}
\begin{equation} 
\begin{aligned}
\Gamma_{0,n_j}^{(j)}&= \frac{4\pi}{\hbar^2}\sum_{k.l}\big\{|M_{jkl}|^2\big(n_j\langle a_ka_k^\dagger\rangle \langle a_la_l^\dagger\rangle +(n_j+1)\langle a_k^\dagger a_k\rangle \langle a_l^\dagger a_l\rangle\big )\delta(\Omega_j-\Omega_k-\Omega_l) \\ & + 2|M_{klj}|^2\big(n_j\langle a_ka_k^\dagger\rangle \langle a_l^\dagger a_l\rangle +(n_j+1)\langle a_k^\dagger a_k\rangle \langle a_l a_l^\dagger\rangle\big )\delta(\Omega_k-\Omega_l-\Omega_j)
\big\},\\
\Gamma_{\downarrow,n_j}^{(j)}&= \frac{4\pi}{\hbar^2}(n_j+1)\sum_{k.l}\big\{|M_{jkl}|^2\langle a_ka_k^\dagger\rangle \langle a_la_l^\dagger\rangle \delta(\Omega_j-\Omega_k-\Omega_l)+ 2|M_{klj}|^2\langle a_ka_k^\dagger\rangle \langle a_l^\dagger a_l\rangle\delta(\Omega_k-\Omega_l-\Omega_j)
\big\},~~{\rm and}\\
\Gamma_{\uparrow,n_j}^{(j)}&= \frac{4\pi}{\hbar^2}n_j\sum_{k.l}\big\{|M_{jkl}|^2\langle a_k^\dagger a_k\rangle \langle a_l^\dagger a_l\rangle \delta(\Omega_j-\Omega_k-\Omega_l)+ 2|M_{klj}|^2\langle a_k^\dagger a_k\rangle \langle a_l a_l^\dagger\rangle\delta(\Omega_k-\Omega_l-\Omega_j)
\big\}.
\end{aligned}
\label{eq12}
\end{equation}
\end{widetext}
The expectation values in Eq. \eqref{eq12} are given by $\langle a_p^\dagger a_p\rangle = \sum_{n_p=0}^\infty \rho_{n_p,n_p}^{(p)} n_p$ and $\langle a_p a_p^\dagger\rangle = \sum_{n_p=0}^\infty \rho_{n_p,n_p}^{(p)}(n_p+1)$. It is straightforward to see that the thermal, i.e. detailed balance governed populations yield a steady state solution of Eq. \eqref{eq5duplicate}, but we want to see under which conditions they would be reached. 

\begin{figure}
		\centering
		\includegraphics [width=\columnwidth] {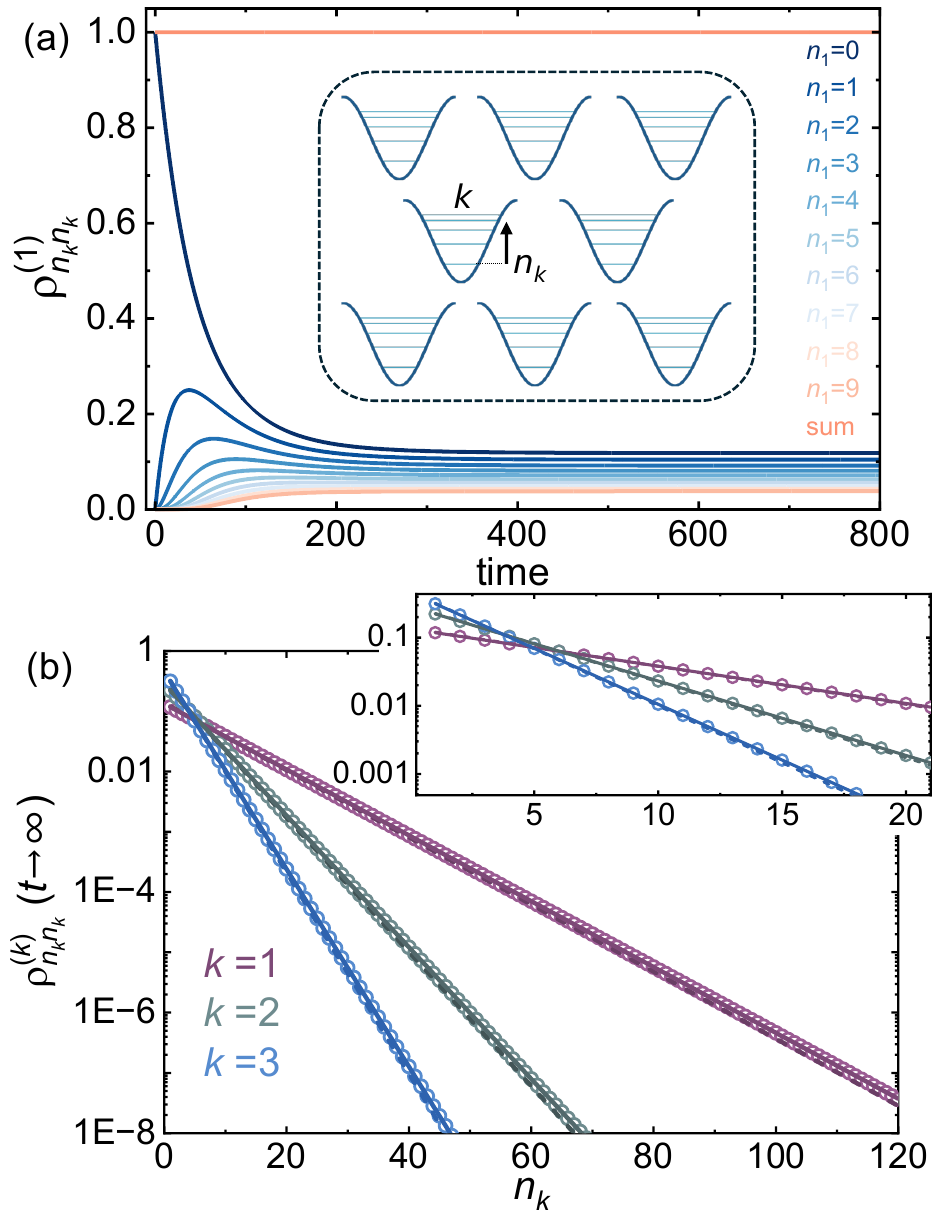}
		\caption{Thermalization of a set of qubits under three-wave mixing in form of weakly anharmonic oscillators (WAOs). (a) Time traces of the lowest energy WAO (sketched in the inset) on different energy levels $n_1 =0,1,..., 9$ calculated numerically. Time $t$ is given in units $1/\Delta\Omega$, and the parameters are $N=100$, $n_k^{\rm{max}}=150$ (truncation of the number of levels included for each WAO),  $|M_{ijk}|^2=0.001$ for all $i,j,k$, and the initial state is $\rho_{10,10}^{(10)}(0)=1$ with all the other WAOs in their ground state. (b) The asymptotic $t\rightarrow \infty$ populations on the level $n_k$ of the WAOs $k=1,2,3$ with parameters as in (a). The numerical results are shown by symbols, whereas the thermal populations of Eq. \eqref{eq15} are shown by lines, based on both evaluating $\tilde \beta$ from Eq. \eqref{eq16} exactly (solid line) or from the analytic approximation Eq. \eqref{eq17} (dashed line). These two results almost coincide with these parameters, and both agree with the numerical long time values
			\label{almostHO-Snail}}
\end{figure}

Numerical results for the system of qubits obeying Eqs. \eqref{eq5duplicate} and \eqref{eq12} are presented in Fig. \ref{almostHO-Snail}. The time traces of the populations on 10 lowest levels for the $k=1$ WAO are shown in (a) with $N=100$ and having uniform distribution of energies such that $\Omega_k=k\Delta\Omega$. In this example the initial state is chosen to be such that the $k=10$ WAO is populated on the level $10$, whereas all the others are in their ground state. For simplicity, we have set all the $|M_{pqr}|^2$ equal, but as shown in panels (a) of Figs. S1 - S3 in Supplemental Material, this choice does not limit our conclusions.

To assess the numerical results against the thermal state, we recall the equilibrium populations on the harmonic oscillators
\begin{equation} \label{eq15}
\rho_{n_p,n_p}^{(p)}=(1-e^{-\beta \hbar\Omega_p})e^{-\beta \hbar\Omega_p  n_p },
\end{equation}
where $\beta=(k_BT)^{-1}$ is the inverse temperature. 
This temperature in equilibrium is determined by energy conservation, i.e. with initial excitation energy $\Delta E =\sum_{k=1}^N \sum_{n_k=0}^\infty n_k \hbar \Omega_k$ in this case. Equating the initial and final energies, we have
\begin{equation} \label{eq16}
\Delta E =\sum_{k=1}^N \frac{\hbar\Omega_k}{e^{\beta \hbar\Omega_k}-1}.
\end{equation}
We obtain the temperature in the thermal state by making an integral approximation of the sum in Eq. \eqref{eq16}, or by solving Eq. \eqref{eq15} exactly numerically. The former method yields in the dimensionless form
\begin{equation} \label{eq17}
\tilde \beta \approx\frac{\pi}{\sqrt{6\Delta \varepsilon}},
\end{equation}
where $\tilde\beta=\beta\hbar\Delta\Omega$ and $\Delta \varepsilon = \Delta E/\hbar\Delta\Omega$. This is a valid approximation for $N,\Delta \varepsilon \gg 1$, but otherwise the numerical evaluation of Eq. \eqref{eq16} is needed. Excellent agreement of the numerically calculated asymptotic long-time distribution in Fig. \ref{almostHO-Snail} (a) with the analytical estimate \eqref{eq17} based on energy conservation in the isolated system is seen in Fig. \ref{almostHO-Snail} (b). See panels (b) of Figs. S1 - S3 in Supplemental Material for a few further examples.  

For completeness and illustration, let us consider an ideal qubit instead of a WAO, i.e. a pure two-level system. We will now consider also the evolution of the off-diagonal elements, i.e. dephasing. For the three-wave mixing in the ideal (two-level) qubits, we use Eqs. \eqref{eq1} and \eqref{eq6a} and obtain the master equation for the ground ($g$) and excited ($e$) states as
\begin{equation}\label{eq25}
	\dot{\rho}^{(j)}_{gg}=-\Gamma^{(j)}_-\rho^{(j)}_{gg}+\Gamma^{(j)}_+\rho^{(j)}_{ee},\,\,\, \dot{\rho}^{(j)}_{ge}=-\frac{1}{2}\Gamma^{(j)}_\Sigma \rho^{(j)}_{ge}, 
\end{equation}
where
\begin{eqnarray} \label{eq7}
&&\Gamma^{(j)}_+ \equiv \frac{4\pi}{\hbar^2} \sum_{k,l}\big{(}|M_{jkl}|^2\rho^{(k)}_{gg}\rho^{(l)}_{gg}\delta(\Omega_j-\Omega_k-\Omega_l)\nonumber\\&&~~~~~+2|M_{klj}|^2\rho^{(k)}_{gg}\rho^{(l)}_{ee}\delta(\Omega_k-\Omega_l-\Omega_j)\big{)},\nonumber\\&&\Gamma^{(j)}_- \equiv \frac{4\pi}{\hbar^2} \sum_{k,l}\big{(}|M_{jkl}|^2\rho^{(k)}_{ee}\rho^{(l)}_{ee}\delta(\Omega_j-\Omega_k-\Omega_l)\nonumber\\&&~~~~~+2|M_{klj}|^2\rho^{(k)}_{ee}\rho^{(l)}_{gg}\delta(\Omega_k-\Omega_l-\Omega_j)\big{)},
\end{eqnarray}
and $\Gamma^{(j)}_\Sigma \equiv \Gamma^{(j)}_+ + \Gamma^{(j)}_-$. Here we have used $\langle a_l a^\dagger _l\rangle=\rho^{(l)}_{gg}$ and $\langle a^\dagger_l a _l\rangle=\rho^{(l)}_{ee}=1-\rho^{(l)}_{gg}$ for the expectation values $\langle \cdot\rangle$ of a two-level system. 
%\cred{As for the WAO's, we discuss quantitatively the dynamics of a qubit in this setup influenced by different coupling mechanisms and regimes. First, one can see immediately, that the steady-state of Eq. \eqref{eq5} for $\rho^{(p)}_{gg}$ is satisfied by thermal populations obeying detailed balance governed populations 
%\begin{equation} \label{eq7b1}
%\rho^{(p)}_{ee}=e^{-\beta \hbar\Omega_p}\rho^{(p)}_{gg}.
%\end{equation}
%This statement is true for both linear coupling and three-wave mixing above. The question is again: can this state be achieved in the long time limit?}
\begin{figure}
		\centering
		\includegraphics [width=\columnwidth] {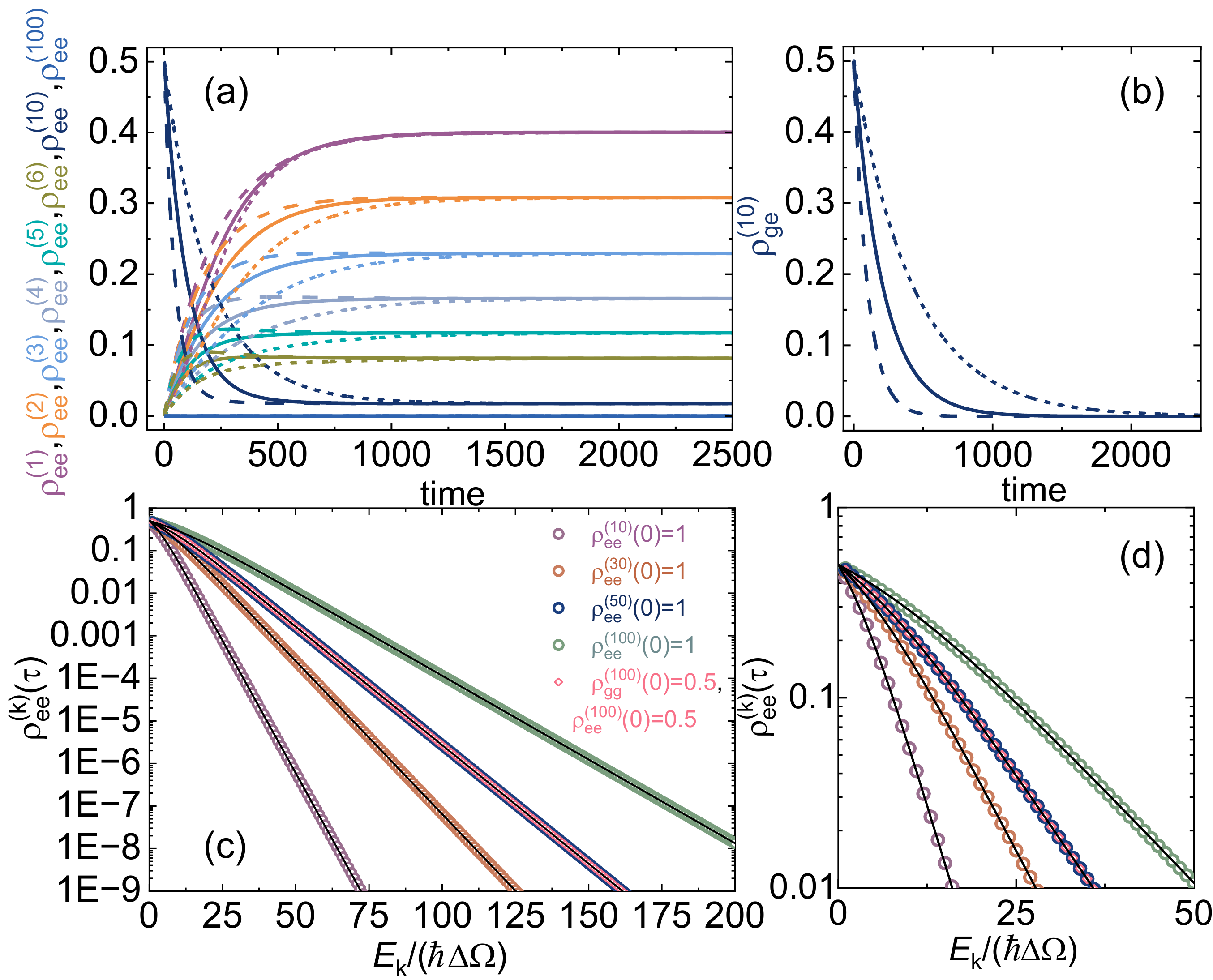}
		\caption{Dynamics and thermalization of $N$ qubits in form of ideal two-level systems under three-wave mixing after initial excitation of a selected qubit with other qubits in the ground state, and with different coupling modalities. (a) The time traces of populations of a few selected qubits presented with different coupling forms: solid curves correspond to $|M_{pqr}|^2=0.001$ for all combinations of $p,q,r$, dashed curves to $|M_{pqr}|^2=0.0002p\,\exp(-|q-r|/N)$, and short-dashed curves (for demonstrating the robustness of thermalization) to a more unrealistic form $|M_{pqr}|^2=(0.005/p)\,\exp(|q-r|/N)$, with initial superposition state $\rho_{gg}^{(10)}(0)=\rho_{ee}^{(10)}(0)=0.5$ and $N=100$. (b) Dynamics of the off-diagonal element $\rho_{ge}^{(10)}$ for three different coupling forms $|M_{pqr}|^2$, with the same parameter values as in (a). (c) Asymptotic in time ($t\rightarrow \infty$) populations $\rho_{ee}^{(k)}(\infty)$ of the $k$:th qubit ($\Omega_k=k\Delta\Omega$) under various initial conditions for $N=300$ qubits calculated numerically presented by circle symbols with different colors. The results are indistinguishable from the thermal populations of two-level systems according to Eq. \eqref{fermi} with temperature obtained from Eq. \eqref{eq10} (solid lines). The coupling strength is $|M_{pqr}|^2 = 0.005$ for all $p,q,r$. The time is displayed in units $1/\Delta \Omega$. (d) Zoom-out of (c) for the lower energy qubit data. 
			\label{rhogg-rhoge}}
\end{figure}

In Fig. \ref{rhogg-rhoge}\,(a) we present results for ideal qubits of a case where $|M_{jkl}|^2$ have three different energy dependences. Furthermore, we give a uniform distribution of the energies $\hbar \Omega_p$ of the $N$ qubits as before. We initiate the simulation with a superposition state of qubit $10$ with $\rho_{gg}^{(10)}(0)=\rho_{ee}^{(10)}(0)=0.5$ and with other qubits in the ground state. The populations stabilize in the long time limit to an $\Omega_k$-dependent common value for all these choices of the couplings. At the same time, the off-diagonal element of the qubit initially in superposition decays exponentially under the same conditions, see Fig. \ref{rhogg-rhoge} (b).

We next find the temperature in the equilibrium state when the system is initialized in a general state $|\psi_k\rangle =c_g^{(k)}|g_k\rangle +c_e^{(k)}|e_k\rangle$ for each qubit. Here $|c_g^{(k)}|^2+|c_e^{(k)}|^2=1$ and $|g_k\rangle,|e_k\rangle$ are the ground and excited states, respectively. This means that initially $\rho^{(k)}_{gg}=1-\rho^{(k)}_{ee}  =|c_g^{(k)}|^2$ and $\rho^{(k)}_{ge} =c_g^{(k)}c_e^{(k)*}$. For the thermal state of a two level system, we expect
\begin{equation} \label{fermi}
    \rho^{(n)}_{ee} = 1/(1+e^{\beta \hbar\Omega_n}).
\end{equation}
The value of $\beta$ is again determined by the energy conservation condition in this case as
\begin{equation} \label{eq9}
\Delta E=\sum_{k=1}^N  \frac{\hbar\Omega_k}{1+e^{\beta \hbar\Omega_k}},
\end{equation}
where $\Delta E = \sum_{k=1}^N |c_e^{(k)}|^2 \hbar\Omega_k$ is the initial energy of the system. Approximating the sum by an integral, we obtain an approximation of $\tilde\beta$ as
\begin{equation} \label{eq10}
\tilde\beta\approx\frac{\pi}{\sqrt{12\Delta\varepsilon}},
\end{equation}
when again assuming that $\Omega_k=k\Delta\Omega$ and $\Delta\varepsilon=\Delta E/\hbar\Delta\Omega$. The populations of each qubit in the long-time limit for the system initialized in different states are shown in Fig. \ref{rhogg-rhoge} (c) and (d). The symbols are from the numerical simulation, whereas the solid lines, fully overlapping the symbols, are drawn using the approximate value of $\tilde\beta$ from Eq. \eqref{eq10}. Perfect thermalization is seen: we have run simulations with various initial states, and the qubit system always reaches the thermal distribution determined only by the initial energy.
\begin{figure}
		\centering
		\includegraphics [width=\columnwidth] {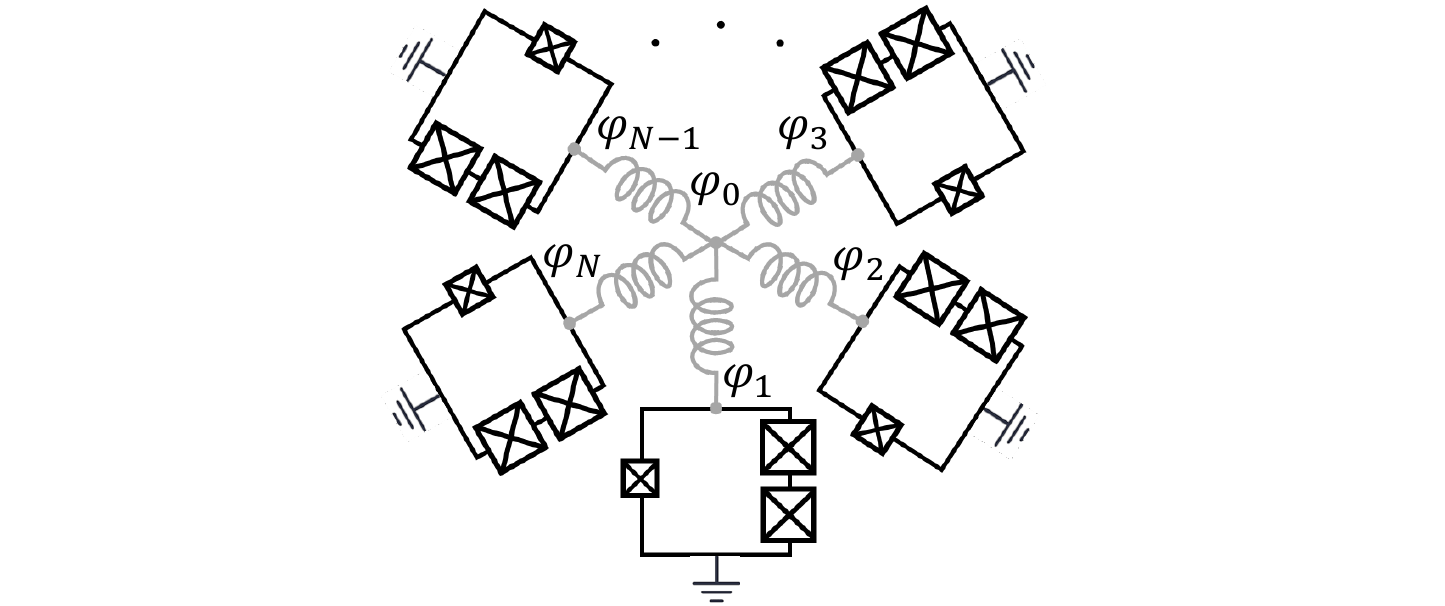}
		\caption{A possible experimental realization: weakly anharmonic oscillators in form of asymmetric Josephson junction loops are connected all-to-all via inductors.
			\label{Snail-expporosal}}
\end{figure}

{\sl Possible experimental realization:} An example of a circuit which obeys the dynamics described above is shown in Fig. \ref{Snail-expporosal}, consisting of asymmetric SQUIDs (superconducting quantum interference devices). They can present three-wave mixing at certain biasing conditions \cite{Frattini2017}, and generally, being composed of Josephson junctions, non-linearity is inherent to them. We consider only third order non-linearity here, and write the Hamiltonian in the form
\begin{equation} \label{exp1}
\mathcal{H} =\sum_{k=1}^N (\frac{\varphi_k^2}{2L_k}+\frac{(\varphi_k-\varphi_0)^2}{2 L^{(\rm c)}_k}+\alpha_k \varphi_k^3)+\mathcal{H}_{\rm C}.
\end{equation}
Here $\varphi_k$ is the phase of SQUID $k$, and $\varphi_0$ the phase of the central node, $L_k$ is the effective inductance of SQUID $k$, $L^{(\rm c)}_k$ is the $k$:th coupling inductance, $\alpha_k$ is the strength of the third order non-linearity of SQUID $k$, and $\mathcal{H}_{\rm C}$ is the charging Hamiltonian given by the capacitances in the circuit. Charge conservation yields for the currents $I_\ell$ through the couplers, $\sum_{\ell=1}^N I_\ell = \sum_{\ell=1}^N (\varphi_\ell -\varphi_0)/L^{(\rm c)}_k =0$. Thus, $\varphi_0=\sum_{\ell=1}^N (\varphi_\ell/L^{(\rm c)}_\ell) /\sum_{\ell=1}^N (1/L^{(\rm c)}_\ell)$. Now $\mathcal{H}$ can be re-written in the form
\begin{equation} \label{exp2}
\mathcal{H} =\sum_{k=1}^N \frac{\varphi_k^2}{2L_k^*}-\sum_{k\neq l}g_{kl}\varphi_k\varphi_l+\sum_{k=1}^N\alpha_k \varphi_k^3+\mathcal{H}_{\rm C},
\end{equation}
where the new parameters $L_k^*$ and $g_{kl}$ are given by the inductances $L_i,L^{(\rm c)}_i$. Next we diagonalize the Josephson part of the Hamiltonian by introducing new phases $\phi_\ell$, such that $\varphi_k =\sum_{\ell=1}^N \lambda_{k\ell}\phi_\ell$. After diagonalization, we have the total Hamiltonian in the form
\begin{equation} \label{exp3}
\mathcal{H} =\sum_{k=1}^N (\frac{\phi_k^2}{2L_k^{**}}+\mathcal{H}_{\rm C})+(\sum_{k,1}\Lambda_{k,l} \phi_k)^3,
\end{equation}
where again the new parameters $L_k^{**},\Lambda_{k,l}$ arise from the inductances in the system. The first part of the Hamiltonian represents $k$ harmonic oscillators, such that $\phi_k \propto (a_k+a_k^\dagger)$. Inserting this dependence in Eq. \eqref{exp3}, we finally obtain the perturbation including the three-wave mixing terms of Eq. \eqref{eq6}.

{\sl Discussion:} We have demonstrated quantitatively, that a system within an ensemble of coupled qubits reaches a thermal state due to non-linearity, in our case three-wave mixing. Excellent agreement with the long-time limit of the numerical solution of the non-linear master equation of the system and the expected thermal state has been observed. Although not proven analytically, our results suggest that the systems reach this thermal state via three-wave mixing irrespective of the particular energy dependence of the coupling strength, at least as long as non-vanishing all-to-all coupling is present. In each case we find the temperature of the final state by a simple energy conservation argument. We have also investigated dephasing in the system of qubits, and found that the off-diagonal elements of the qubits' density matrix vanish exponentially in time, as expected. Our model, although applying to realistic quantum systems, has certain limitations. First, it applies in weak coupling limit, where energy conservation yields strict delta-function conditions for the energies of the interacting elements. Yet, compromising this requirement of extreme weak coupling does not change the dynamics qualitatively, see, e.g. \cite{Pekola2023}. The question of thermalization vs. many body localization in three-wave mixing was discussed recently in \cite{Ankerhold2026}, where localization is ascribed to fragmentation of the Hilbert space determined by non-uniformity of the energy distribution. Related to this issue, we believe that assuming uniform distribution of energies (equidistant for a finite system) of the participating elements in our work does not limit the value of the conclusions, in particular in the large $N$ limit, when the density of energies is much higher than the inverse coupling strength smearing the energy levels. 

	{\it Acknowledgments ---} We thank Joachim Ankerhold, Andrew Cleland, Paolo Muratore-Ginanneschi, Xuntao Wu, and Ilari M\"akinen for useful discussions. This work has received funding from the European Union’s Research and Innovation Programme, Horizon Europe, under the Marie Sklodowska-Curie Grant Agreement No. 101150440 (TcQTD). we acknowledge funding from the Research Council of Finland Centre of Excellence programme grant 336810 and grant 349601 (THEPOW) and QuantERA II Programme that has received funding from the EU’s H2020 research and innovation programme under the GA No 101017733. \\
    	For correspondence:\\ \texttt{email:jukka.pekola@aalto.fi}\\ \texttt{email:bayan.karimi@aalto.fi}

\clearpage

\section{End matter}
\begin{figure} [h!]
		\centering
		\includegraphics [width=\columnwidth] {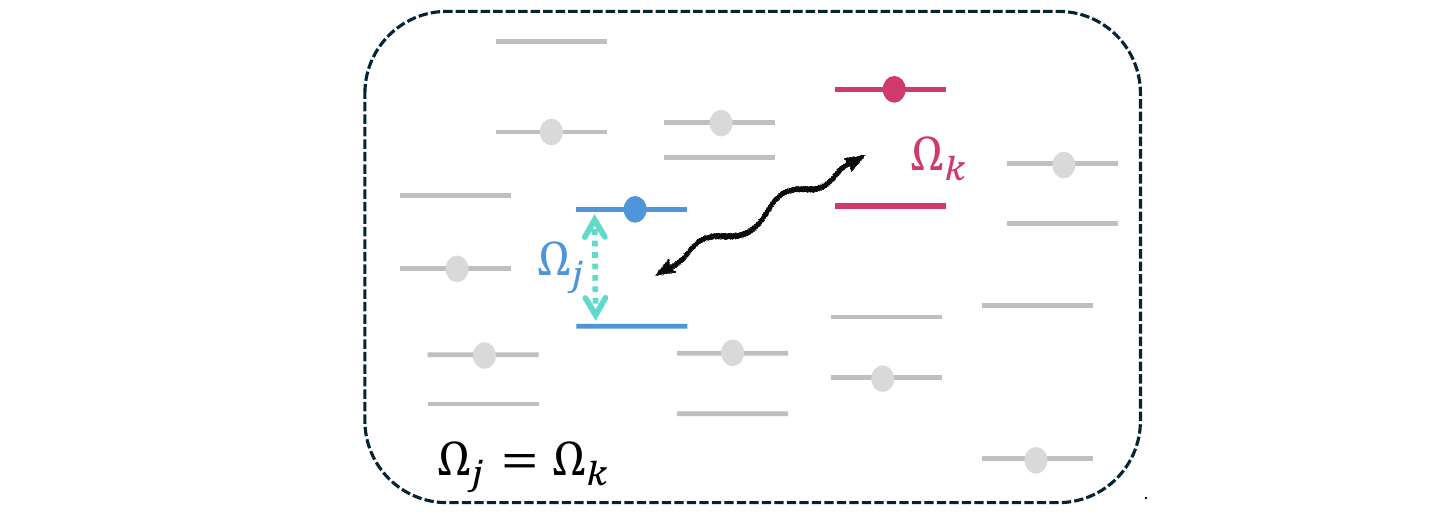}
		\caption{Linear coupling in all-to-all coupled qubits. Here qubits interact efficiently only when they are degenerate.
			\label{lincoupling}}
\end{figure}
{\sl Linear coupling:} For linear coupling (Fig.~\ref{lincoupling}), which is the simplest description of the system, for instance for capacitively or inductively coupled harmonic oscillators, we write in the rotating wave approximation
\begin{equation}\label{eq2}	\hat{\mathbb{V}}^{(j)}=\sum_{k\neq j}\big{(}g_{jk}\hat{a}_j^\dagger \hat{a}_k +g_{jk}^* \hat{a}_k^\dagger \hat{a}_j \big{)}.
\end{equation}
Here $g_{jk}$ is the coupling constant between WAOs $j$ and $k$. By standard manipulations we can combine Eqs. \eqref{eq1} and \eqref{eq2} and obtain the master equation \eqref{eq5duplicate} for the diagonal elements of $\rho^{(j)}$ on the level $n_j$ in the harmonic oscillator approximation. In this case,
%\begin{eqnarray} \label{eq6}
$\Gamma_{0,n_j}^{(j)}= \frac{2\pi}{\hbar^2}\sum_{k}|g_{jk}|^2\{n_j\langle a_ka_k^\dagger\rangle  +(n_j+1)\langle a_k^\dagger a_k\rangle \}\delta(\Omega_j-\Omega_k)$,
%\end{eqnarray}
%\begin{eqnarray} \label{eq7}
$\Gamma_{\downarrow,n_j}^{(j)}= \frac{2\pi}{\hbar^2}(n_j+1)\sum_{k}|g_{jk}|^2 \langle a_ka_k^\dagger\rangle\delta(\Omega_j-\Omega_k)$,
%\end{eqnarray}
and
%\begin{eqnarray} \label{eq8}
$\Gamma_{\uparrow,n_j}^{(j)}= \frac{2\pi}{\hbar^2}n_j\sum_{k}|g_{jk}|^2 \langle a_k^\dagger a_k\rangle\delta(\Omega_j-\Omega_k)$.
%\end{eqnarray}
The expectation values are given as before for the harmonic oscillators. These equations show immediately that within this model, due to strict $\delta$-function like energy conservation in the transitions, the qubits interact only with degenerate ones, and no thermalization occurs. Later we will see that this statement is true even when coupling is stronger leading to broadening of the energy levels. 
%We also defer the discussion of off-diagonal elements to the case of qubits being pure two level systems.

For completeness we consider an ideal qubit instead of a WAO, now in linear coupling \eqref{eq2}. 
We then obtain again the master equation \eqref{eq25} but with
\begin{equation} \label{eq4b}
\Gamma^{(j)}_+ \equiv \frac{2\pi}{\hbar^2} \sum_{k\neq j}|g_{jk}|^2\rho^{(k)}_{gg}\delta(\Omega_j-\Omega_k),
\end{equation}
\begin{equation} \label{eq4c}
\Gamma^{(j)}_- \equiv \frac{2\pi}{\hbar^2} \sum_{k\neq j}|g_{jk}|^2\rho^{(k)}_{ee}\delta(\Omega_j-\Omega_k),
\end{equation}
and $\Gamma^{(j)}_\Sigma \equiv \Gamma^{(j)}_+ + \Gamma^{(j)}_-$.
We have ignored the phase shift in Eq. \eqref{eq25} (for $\rho^{(j)}_{ge}$) originating from the Cauchy principal value, see Section I in the Supplemental Material for the detailed derivation of it for this setup.

{\sl Broadening of the levels:} In the weakly and linearly coupled system, the given qubit interacts only with the degenerate ones according to the energy conservation implied by Eqs. \eqref{eq4b} and \eqref{eq4c}. This means that strictly only in the limit $N\rightarrow \infty$ one can observe relaxation type dynamics \cite{Fonda,Khalfin}, and only among the degenerate ones. Yet this linearly coupled two-level case allows us to analyze the problem also beyond the extreme weak-coupling regime \cite{Pekola2023,Entropypekola2024,blueprint2026}, e.g. in the case of a single-excitation in the system~\cite{Lee}. This is because the Hilbert space remains small due to the conservation of number of excitations. When the coupling increases, the qubit interacts with those surrounding ones whose energies are within about the overall relaxation rate of the qubit in question. As an example, consider a single-excitation initial state of a system with $N$ qubits. The increased coupling is reflected by the long time Lorentzian population distribution of the surrounding qubits around the energy of the central qubit \cite{Pekola2023}, instead of delta-function distribution in the extreme weak coupling. Thus, the system finds a long-time non-thermalized steady state, where the population of the initially excited qubit also follows the distribution of the $N-1$ other qubits. The main conclusion is that, even in the long time limit, the qubits with frequency away from the central one remain untouched, i.e. the detailed balance governed populations 
\begin{equation} \label{eq7b}
\rho^{(p)}_{ee}=e^{-\beta \hbar\Omega_p}\rho^{(p)}_{gg} 
\end{equation}
cannot be reached even in this case, although they present a steady-state solution of the master equation.
%The above analysis demonstrates again exact energy conservation in the weak coupling limit witnessed by the delta-functions in the transition rates in Eqs. \eqref{eq4b} and \eqref{eq4c}. Now we can address the question whether the non-thermalization in the linear coupling case would be a special feature of extreme weak coupling, where only the qubits with strictly equal energy to the excited one get populated. Yet this conclusion is more general. In \cite{Pekola2023} the present case of linear coupling was analyzed in the situation where one qubit is initially excited, and coupled to the rest of them via $g_{0j}$. That problem can be solved exactly for arbitrary coupling value, with the result that the $N$ qubits have in the long-time limit populations $ \rho_{ee}^{(j)}$ given by the Lorentzian centered around the frequency $\Omega$ of the initially excited qubit as
Quantitatively, the long-time distribution reads~\cite{Pekola2023}
\begin{equation} \label{eq5a}
\rho_{ee}^{(j)} = \frac{1}{\pi (N-1)}\frac{\Delta \Omega}{\Omega}\frac{g_{0j}^2}{\langle g^2\rangle}\frac{\Gamma_0/(2\Omega)}{(\Gamma_0/(2\Omega))^2+(\Omega_j/\Omega -1)^2}.
\end{equation}
Here, $g_{0j}$ is the coupling of the central qubit to the $j$:th external one, $\langle g^2\rangle =\sum_{j=1}^{N-1} g_{0j}^2 /(N-1)$ is the average of the squared couplings, $\Delta \Omega/\Omega$ is the uniform (normalized) spread of qubit frequencies symmetrically around that of the central qubit, $\Omega$, and $\Gamma_0 = \frac{2\pi}{\hbar^2}\frac{N-1}{\Delta \Omega} \langle g^2\rangle$ is the overall relaxation rate related to the coupling strength. %Numerically calculated examples are shown in Fig. \ref{fig2} for a few sets of parameters. 
The weak coupling limit $\Gamma_0 \rightarrow 0$ yields the expected delta-function governed energy conservation. See Fig. S4 of the Supplemental Material for further information.

It is, however, illustrative to see that for a large set of qubits, $N\rightarrow \infty$, even in the linear and weak-couling regime, the subsystem formed of qubit $j$ relaxes and dephases, as given by Eqs. \eqref{eq25}, \eqref{eq4b} and \eqref{eq4c}: both the excited state population and the off-diagonal element of its density matrix decay (almost) exponentially in time with the rates $\Gamma^{(j)}_\Sigma$ and $\Gamma^{(j)}_\Sigma /2$, respectively. In this limit $\Gamma^{(j)}_\Sigma = \frac{2\pi}{\hbar^2}\nu_0 \langle g^2\rangle$, where $\nu_0$ is the density of qubit (angular) frequencies at $\Omega_j$. The residual populations of the qubits in the long time limit are, however, not thermal even for large $N$.

\clearpage
\onecolumngrid
\setcounter{section}{0}
\renewcommand{\thesection}{S\arabic{section}}
\setcounter{figure}{0} % Resets the figure counter to 0
\renewcommand{\thefigure}{S\arabic{figure}} % Changes the figure numbering to S1, S2, etc.
\renewcommand{\figurename}{Fig.} % Ensures the caption starts with "Fig." instead of "Figure"
\setcounter{table}{0}
\renewcommand{\thetable}{S\arabic{table}}%
\setcounter{equation}{0} % Reset equation numbering
\renewcommand{\theequation}{S\arabic{equation}} % Prefix with S. and Arabic number
\section*{Supplemental Material:}
	\section{Dephasing of the two-level systems}
Here we take a closer look at the evolution of the off-diagonal element of the density matrix of the $j$:th qubit. 
Equation (4) in the main text arises by ignoring the Cauchy principal value of the integral in
\begin{equation}\label{e1}
	\dot{\rho}^{(j)}_{ge}(t)=-\frac{1}{\hbar^2}\sum_{k}|g_{jk}|^2  \int_{-\infty}^t d\tau e^{i(\Omega_k-\Omega_j)(\tau-t)}\rho^{(j)}_{ge}(t) . 
\end{equation}
With the help of the identity
\begin{equation}\label{e2}
	\int_{-\infty}^0 d\tau e^{i\lambda \tau} = \pi \delta(\lambda) -i \lim_{\epsilon \rightarrow 0} \frac{\lambda}{\epsilon^2 +\lambda^2}  
\end{equation}
we obtain
\begin{equation}\label{e3}
	\dot{\rho}^{(j)}_{ge}(t)=-\frac{1}{\hbar^2}\sum_{k}|g_{jk}|^2  \big(\pi \delta(\Omega_k-\Omega_j) -i \lim_{\epsilon \rightarrow 0} \frac{\Omega_k-\Omega_j}{\epsilon^2 +(\Omega_k-\Omega_j)^2}\big)\rho^{(j)}_{ge}(t) . 
\end{equation}
Now, assuming uniform uncorrelated distributions of frequencies and couplings in the large $N$ limit, we obtain
\begin{equation}\label{e4}
	\dot{\rho}^{(j)}_{ge}(t)=-\big(\frac{\pi}{\hbar^2}\sum_{k}|g_{jk}|^2  \delta(\Omega_k-\Omega_j)-i \nu_j \langle g_j^2\rangle \lim_{\epsilon \rightarrow 0} \int_{\Omega_j-\Delta^{(j)}_1}^{\Omega_j+\Delta^{(j)}_2}\frac{\Omega-\Omega_j}{\epsilon^2 +(\Omega-\Omega_j)^2}\big)\rho^{(j)}_{ge}(t) . 
\end{equation}
Here, the distribution of qubit frequencies spans over $[\Omega_j-\Delta^{(j)}_1,\Omega_j+\Delta^{(j)}_2]$, with the density $\nu_j=(N-1)/(\Delta^{(j)}_1+\Delta^{(j)}_2)$ and the average of couplings is taken as $\langle g_j^2\rangle =\sum_k g_{jk}^2/(N-1)$. We can then write Eq. \eqref{e4} into
\begin{equation}\label{e5}
	\dot{\rho}^{(j)}_{ge}(t)=-(\Gamma_\Sigma - i P_j) \rho^{(j)}_{ge}(t), 
\end{equation}
where $\Gamma_\Sigma^{(j)}=\frac{2\pi}{\hbar^2}\sum_{k}|g_{jk}|^2  \delta(\Omega_k-\Omega_j)$ as in the main text, and $P_j\equiv\frac{2}{\hbar^2}\nu_j \langle g_j^2\rangle\ln(\Delta^{(j)}_2/\Delta^{(j)}_1)$.

Equation \eqref{e5} gives then
\begin{equation}\label{e6}
	\rho^{(j)}_{ge}(t)=  \rho^{(j)}_{ge}(0)e^{i P_j t} e^{-\Gamma_\Sigma^{(j)} t/2}.
\end{equation}
We see that the Cauchy principal value yields the same decay rate of the magnitude of the off-diagonal elements ($\Gamma_\Sigma^{(j)} /2$) as discussed in the main text; it only gives an additional phase factor $e^{i P_j t}$. In the special case of a symmetric distribution, $\Delta^{(j)}_2=\Delta^{(j)}_1$, $P_j$ vanishes and we recover the standard expression of the main text.

\begin{figure}%[h!]
		%\centering
		\includegraphics [width=\columnwidth] {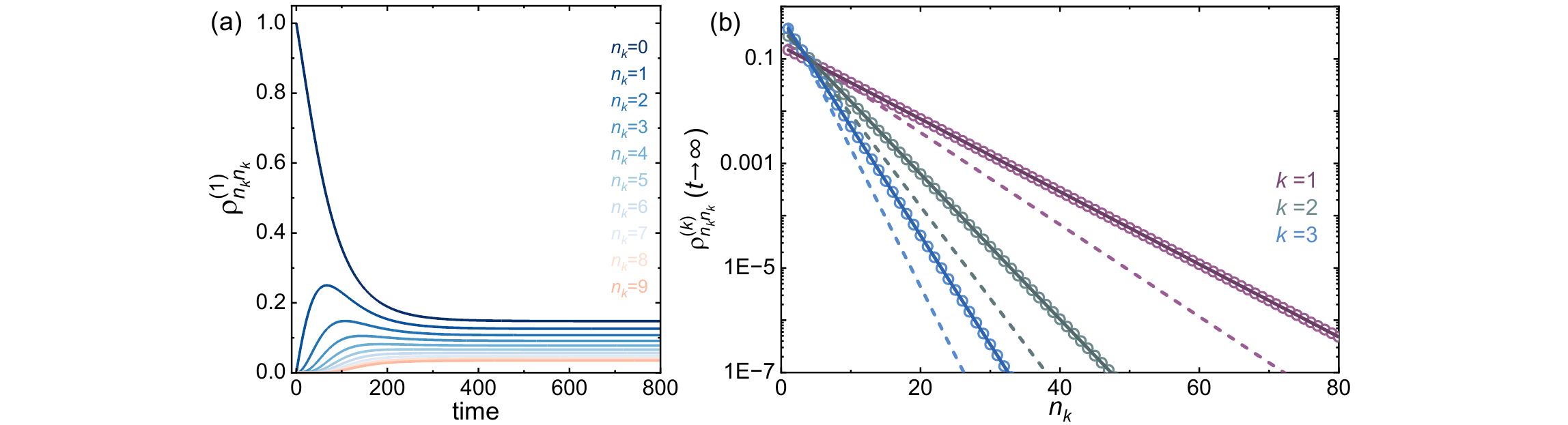}
		\caption{Thermalization of a set of coupled weakly anharmonic oscillators (WAOs). (a) Time traces of the lowest energy WAO (sketched in the inset) on different energy levels $n_0 =0,1,..., 9$ calculated numerically. Time $t$ is given in units $1/\Delta\Omega$, and the parameters are $N=10$, $n_k^{\rm{max}}=150$ (truncation of the number of levels included for each WAO),  $|M_{ijk}|^2=0.001$ for all $i,j,k$, and the initial state is $\rho_{4,4}^{(10)}(0)=1$ with all the other WAOs in their ground state. (b) The asymptotic $t\rightarrow \infty$ populations on the level $n_k$ of the WAOs $k=1,2,3$ with parameters as in (a). The numerical results are shown by symbols, whereas the thermal populations of harmonic oscillators are shown by lines, based on both evaluating $\tilde \beta$ exactly (solid line) or from the analytic approximation of the main text (dashed line). This last expression fails quantitatively in this case due to small $N$. 
			\label{almostHO-Snail-SM-N10}}
\end{figure}

\begin{figure}%[h!]
		%\centering
		\includegraphics [width=\columnwidth] {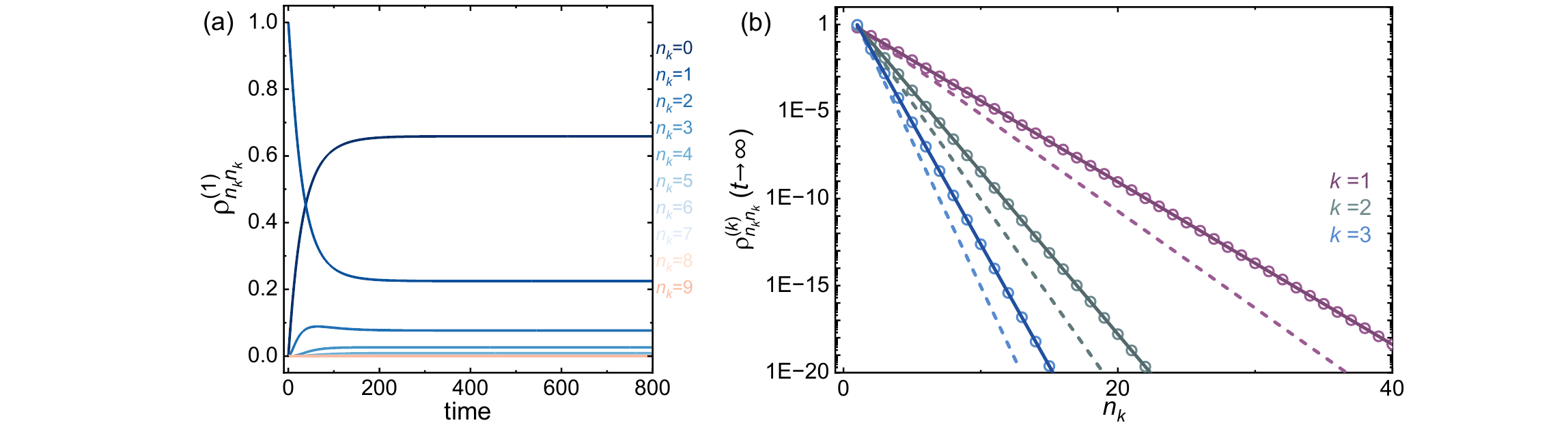}
		\caption{Similar data as in Fig. \ref{almostHO-Snail-SM-N10}, but with parameters $N=10$, $n_k^{\rm{max}}=50$,  $|M_{ijk}|^2=0.01$, and $\rho_{1,1}^{(1)}(0)=1$.
        {almostHO-Snail-SM-N10}
			\label{almostHO-Snail-SM-N10-50states}}
\end{figure}

\begin{figure}%[h!]
		%\centering
		\includegraphics [width=\columnwidth] {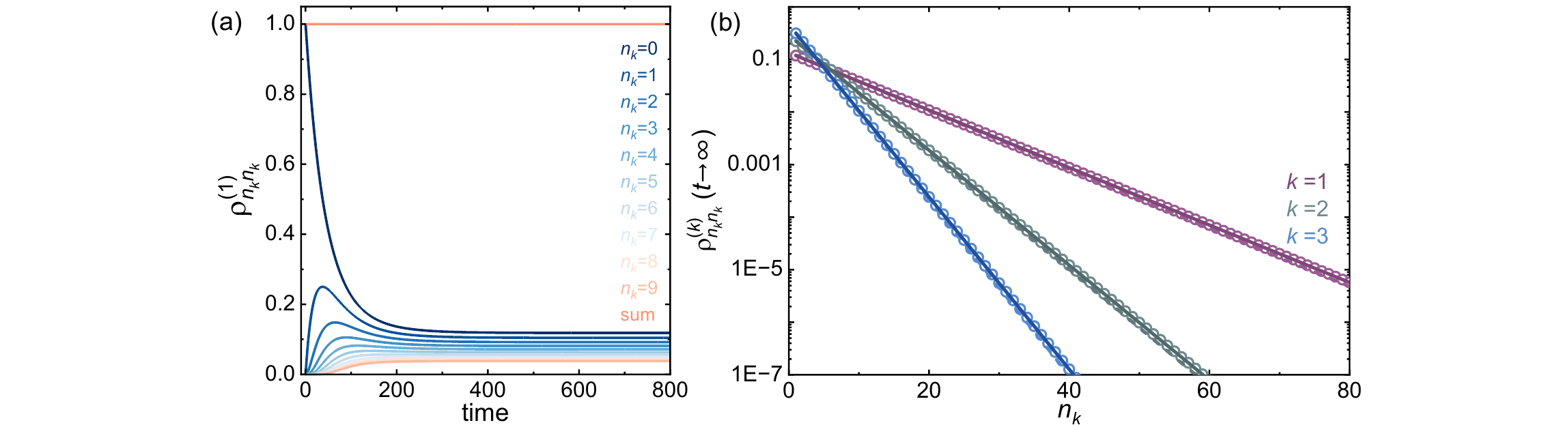}
		\caption{Similar data as in Fig.~\ref{almostHO-Snail-SM-N10}, but with parameters $N=100$, $n_k^{\rm{max}}=150$, $|M_{pqr}|^2=0.001p\,\exp(-|q-r|/N)$, $\rho_{10,10}^{(10)}(0)=1$. Here for large $N$ the analytic expression gives a good approximation. The red horizontal line in (a) presents the sum of all the populations of the system.
			\label{almostHO-Snail-SM-N100-expdiffMijk}}
\end{figure}

%\begin{figure}%[h!]
		%\centering
%		\includegraphics [width=\columnwidth] {rhogg-rhoge-same Mijk-1.pdf}
	%	\caption{Populations of the qubits in (a) and their off-diagonal elements in (b) of the density matrix as a function of time after the initialization under various conditions. Irrespective of the coupling parameters the qubits relax asymptotically to a population determined by the initial energy within the system of qubits, and the off-diagonal elements vanish in this limit. The parameters are $N=100$, $|M_{ijk}|^2=0.005$ (equal for all combinations of $i,j,k$. Solid lines in (a) correspond to the initial condition $\rho_{ee}^{(1)}=1$, dashed lines in (a) to the initial conditions $\rho_{ee}^{(1)}=1$ and $\rho_{ee}^{(2)}=1$, and dash-dotted lines in (a) and (b) to the case where qubit 6 is initialized in a superposition state, $\rho_{gg}^{(6)}=0.5$ and $\rho_{ee}^{(6)}=0.5$. The other qubits are initially in their ground state in each case.
%			\label{rhogg-rhoge-sameMijk}}
%\end{figure}

\begin{figure}%[h!]
		\centering
		\includegraphics [width=0.5\columnwidth] {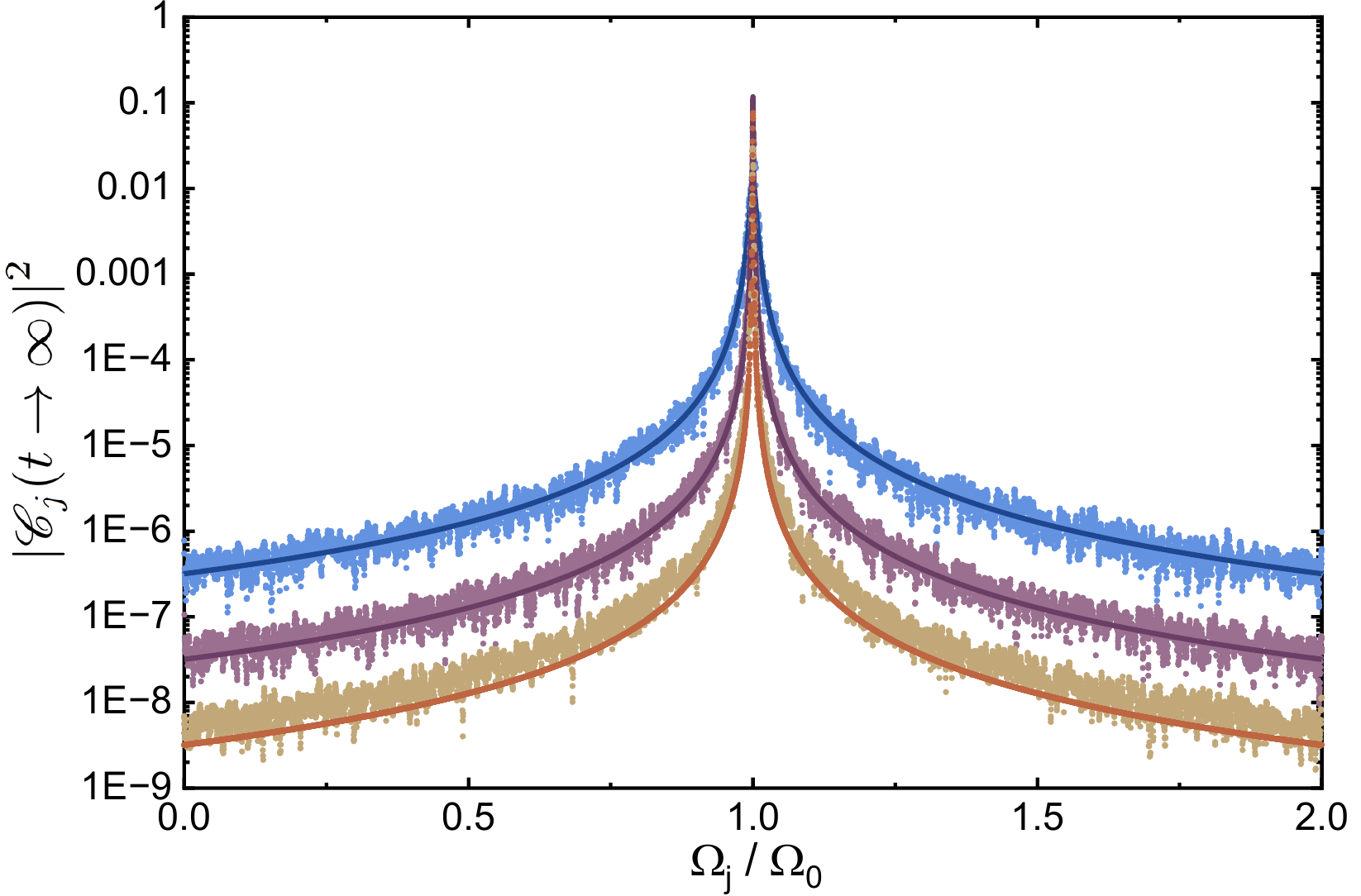}
		\caption{Long-time asymptotic populations of $N$ qubits linearly coupled to the "central one" with energy $\hbar\Omega$ following its initialization the excited state; the other qubits are coupled to the excited one and their energies are uniformly and symmetrically distributed around it. The numerical results are presented for $N=10000$ and for three coupling strengths $\Gamma_0 = 10^{-4}$ (brown), $10^{-3}$ (purple) and $10^{-2}$ (blue) with uniform distribution of couplings. The numerical results as a function of the frequency of each qubit follow a Lorentzian distribution around $\Omega$. The results are compared to the average from the expression given in the main text setting all the couplings equal (solid lines). The numerical populations are correspondingly averaged over 10 neighboring ones in frequency. 
			\label{linearly coupled-lorentzian}}
\end{figure}	

\begin{thebibliography}{99}
    
    \bibitem{Ueda2018} Takashi Mori, Tatsuhiko N. Ikeda, Eriko Kaminishi and Masahito Ueda, Thermalization and prethermalization in isolated quantum systems: a theoretical overview, \href{https://iopscience.iop.org/article/10.1088/1361-6455/aabcdf}{J. Phys. B: At. Mol. Opt. Phys. {\bf 51}, 112001 (2018).}

    \bibitem{Abanin2019} D. A. Abanin, E. Altman, I. Bloch, and M. Serbyn, Many-body localization, thermalization, and entanglement, \href{https://doi.org/10.1103/RevModPhys.91.021001}{Rev. Mod. Phys. {\bf 91}, 021001 (2019).}

    \bibitem{Rigol2016}  L. D'Alessio, Y. Kafri, A. Polkovnikov, and M. Rigol, From quantum chaos and eigenstate thermalization to statistical mechanics and thermodynamics, \href{https://doi.org/10.1080/00018732.2016.1198134}{Adv. Phys. {\bf 65}, 239 (2016).}

    \bibitem{Nandkishore2015} R. Nandkishore and D. A. Huse, Many-body localization and thermalization in quantum statistical mechanics, \href{https://doi.org/10.1146/annurev-conmatphys-031214-014726}{Annu. Rev. Condens. Matter Phys. {\bf 6}, 15 (2015).}

     \bibitem{Andersen2025} T. I. Andersen et al., Thermalization and criticality on an analogue-digital quantum simulator, \href{https://doi.org/10.1038/s41586-024-08460-3}{Nature (London) 638, 79 (2025).}

     \bibitem{Guo2026} Xue-Yi Guo, Energy relaxation via quantum thermalization: A superconducting qubit coupled to an interacting many-body two-level system, APS Open Sci. {\bf 1}, 000099 (2026).

    \bibitem{Weiss2006} T. Kinoshita, T. Wenger, and D. S. Weiss, A quantum Newton's cradle, \href{https://doi.org/10.1038/nature04693}{Nature (London) {\bf 440}, 900 (2006).}

    \bibitem{Schmiedmayer2012} M. Gring, M. Kuhnert, T. Langen, T. Kitagawa, B. Rauer, M. Schreitl, I. Mazets, D. Adu Smith, E. Demler, and J. Schmiedmayer, Relaxation and prethermalization in an isolated quantum system, \href{https://www.science.org/doi/10.1126/science.1224953}{Science {\bf 337}, 1318 (2012).}

    \bibitem{Kaufman2016} A. M. Kaufman, M. E. Tai, A. Lukin, M. Rispoli, R. Schittko, P. M. Preiss, M. Greiner, Quantum thermalization through entanglement in an isolated many-body system, \href{https://www.science.org/doi/10.1126/science.aaf6725}{Science {\bf 353}, 794 (2016).}

    \bibitem{Neill2016}  C. Neill et al., Ergodic dynamics and thermalization in an isolated quantum system, \href{https://www.nature.com/articles/nphys3830}{Nat. Phys. {\bf 12}, 1037 (2016).}

    \bibitem{Chen2021}  F. Chen et al., Observation of strong and weak thermalization in a superconducting quantum processor, \href{https://doi.org/10.1103/PhysRevLett.127.020602}{Phys. Rev. Lett. {\bf 127}, 020602 (2021).}

    \bibitem{Higginbotham2026} Anton V. Bubis, Lucia Vigliotti, Maksym Serbyn, and Andrew P. Higginbotham, Non-equilibrium plasmon liquid in a Josephson junction chain, \href{ Non-equilibrium plasmon liquid in a Josephson junction chain | Science Advances}{Sci. Adv. {\bf 12}, eady7222 (2026).}

    \bibitem{XuntaoWu} Xuntao Wu et al., Modular Quantum Processor with an All-to-All Reconfigurable Router, \href{ https://doi.org/10.1103/PhysRevX.14.041030}{Phys. Rev. X {\ bf 14}, 041030 (2024).}

    \bibitem{Koch2007} Jens Koch, Terri M. Yu, Jay Gambetta, A. A. Houck, D. I. Schuster, J. Majer, Alexandre Blais, M. H. Devoret, S. M. Girvin, and R. J. Schoelkopf, Charge-insensitive qubit design derived from the Cooper pair box, Phys. Rev. A {\bf 76}, 042319 (2007).
    
    \bibitem{Roch2012} N. Roch, E. Flurin, F. Nguyen, P. Morfin, P. Campagne-Ibarcq, M. H. Devoret, and B. Huard, Widely tunable, nondegenerate three-wave mixing microwave device operating near the quantum limit, \href{https://doi.org/10.1103/PhysRevLett.108.147701}{Phys. Rev. Lett. {\bf 108}, 147701 (2012).}
        
    \bibitem{Frattini2017} N. E. Frattini, U. Vool, S. Shankar, A. Narla; K. M. Sliwa, M. H. Devoret, 3-wave mixing Josephson dipole element, \href{https://doi.org/10.1063/1.4984142}{Appl. Phys. Lett. {\bf 110}, 222603 (2017).}

    \bibitem{Ganesan2017} A. Ganesan, C. Do, and A. Seshia, Phononic frequency comb via intrinsic three-wave mixing, \href{https://doi.org/10.1103/PhysRevLett.118.033903}{Phys. Rev. Lett. {\bf 118}, 033903 (2017).}

    \bibitem{Visa2018} Slawomir Simbierowicz, Visa Vesterinen, Leif Gr\"onberg, Janne Lehtinen, Mika Prunnila and Juha Hassel, A flux-driven Josephson parametric amplifier for sub-GHz frequencies fabricated with side-wall passivated spacer junction technology, \href{https://iopscience.iop.org/article/10.1088/1361-6668/aad4f2}{Supercond. Sci. Technol. {\bf 31}, 105001 (2018).}

    \bibitem{Roch2022} A. Ranadive, M. Esposito, L. Planat, E. Bonet, C. Naud, O. Buisson, W. Guichard, and N. Roch, Kerr reversal in Josephson meta-material and traveling wave parametric amplification, \href{https://doi.org/10.1038/s41467-022-29375-5}{Nat. Commun. {\bf 13}, 1737 (2022).}

    \bibitem{Pla2022} D. J. Parker, M. Savytskyi, W. Vine, A. Laucht, T. Duty, A. Morello, A. L. Grimsmo, and J. J. Pla, Degenerate parametric amplification via three-wave mixing using kinetic inductance, \href{https://doi.org/10.1103/PhysRevApplied.17.034064}{Phys. Rev. Appl. {\bf 17}, 034064 (2022).}

    \bibitem{Manucharyan2023} N. Mehta, R. Kuzmin, C. Ciuti, and V. E. Manucharyan, Down-conversion of a single photon as a probe of many-body localization, \href{https://doi.org/10.1038/s41586-022-05615-y}{Nature (London) {\bf 613}, 650 (2023).}

    \bibitem{Buccheri2026} Vittorio Buccheri, Ivo P. C. Cools, Nermin Trnjanin, Ankit Khola, Oleg Shvetsov, Thomas Kanne, Jesper Nygård, Attila Geresdi, and Simone Gasparinetti, Kerr nonlinearity and three-wave mixing in superconducting resonators hosting Al-InAs weak links, \href{https://arxiv.org/abs/2608.28428} {arXiv:2608.28428}.

    \bibitem{3wave-PK} Jukka P. Pekola and Bayan Karimi, Quantum thermalization via multiwave mixing, \href{https://doi.org/10.1103/PhysRevResearch.6.L042023}{Phys. Rev. Research {\bf 6}, L042023 (2024).}

    \bibitem{Ankerhold2026} Evangelos Varvelis, Miriam Resch, and Joachim Ankerhold, Unconventional Thermalization of a Three-Wave-Mixing Model, \href{https://arxiv.org/abs/2607.27028} {arXiv:2607.27028}.

    \bibitem{Fonda} L. Fonda, G. C. Ghirardi and A. Rimini, Decay theory of unstable quantum systems, \href{ https://iopscience.iop.org/article/10.1088/0034-4885/41/4/003}{Rep. Prog. Phys. {\bf 41}, 587 (1978).}
	
	\bibitem{Khalfin} L. A. Khalfin, On the theory of the decay of a quasi-stationary state, \href{http://mi.mathnet.ru/eng/dan22167}{Dokl. Akad. Nauk SSSR {\bf 115:2}, 277 (1957)};  J. Exp. Theor. Phys. {\bf 6}, 1053 (1958).

    \bibitem{Pekola2023} Jukka P. Pekola, Bayan Karimi, Marco Cattaneo, and Sabrina Maniscalco, Long-Time Relaxation of a Finite Spin Bath Linearly Coupled to a Qubit, \href{https://doi.org/10.1142/S1230161223500099}{Open Syst. Inf. Dyn. {\bf 30}, 2350009 (2023).}

    \bibitem{Entropypekola2024} J. P. Pekola and B. Karimi, Heat Bath in a Quantum Circuit, \href{ https://doi.org/10.3390/e26050429}{Entropy {\bf 26}(5), 429 (2024).} 

    \bibitem{blueprint2026} Bayan Karimi, Xuntao Wu, Andrew N. Cleland, and Jukka P. Pekola, Blueprint for experiments exploring the quantum recurrence theorem on a coupled multiqubit system, \href{https://doi.org/10.1103/wv9h-x14y}{Phys. Rev. Research {\bf 8}, L012062 (2026).}

    \bibitem{Lee} T. D. Lee, Some Special Examples in Renormalizable Field Theory, \href{https://journals.aps.org/pr/abstract/10.1103/PhysRev.95.1329}{Phys. Rev. {\bf 95}, 1329 (1954).} 

\end{thebibliography}
	\end{document}